\documentclass[runningheads]{llncs}

\usepackage{amsmath, amssymb, amsfonts}
\usepackage{algorithmic}
\usepackage[ruled,vlined]{algorithm2e}
\usepackage{tabularx}
\usepackage{svg}

\usepackage{graphicx}
\usepackage{array}
\usepackage[caption=false,font=normalsize,labelfont=sf,textfont=sf]{subfig}
\usepackage[dvipsnames]{xcolor}
\usepackage[dvipsnames,table]{xcolor}
\usepackage{tcolorbox}
\usepackage{booktabs} 
\usepackage{array}    
\usepackage{siunitx}  
\usepackage{multirow}  

\usepackage[utf8]{inputenc}
\usepackage{soul}
\usepackage{textcomp}
\usepackage{verbatim}
\usepackage{lipsum}  
\usepackage{setspace}
\usepackage{threeparttable}
\usepackage{marvosym}

\usepackage{hyperref}
\usepackage{stfloats}
\usepackage{balance}
\usepackage{float}

\usepackage{cite} 
\usepackage[T1]{fontenc}

\usepackage{url}
\usepackage{hyperref}
\hypersetup{
    colorlinks=false,     
    linkbordercolor=red,   
    citebordercolor=green, 
    urlbordercolor=blue,   
    pdfborder={0 0 1},      
    breaklinks=true
}
\def\BibTeX{{\rm B\kern-.05em{\sc i\kern-.025em b}\kern-.08em
    T\kern-.1667em\lower.7ex\hbox{E}\kern-.125emX}}

\newcommand{\shortmethod}{'HDI' }
\newcommand{\attackname}{Hop-Decayed Influence }
\newcommand{\attacksurface}{Auxiliary Schema-Level Entity}

\newcommand{\RQone}{Are Auxiliary Schema-Level Entities exploitable, and does \attackname~
effectively compromise GraphRAG systems?}
\newcommand{\RQtwo}{How efficient and scalable is \attackname~compared to 
conventional poisoning approaches?}
\newcommand{\RQthree}{Is \attackname~stealthy enough to evade existing 
GraphRAG defense mechanisms?}

\begin{document}

\title{Hop-Decayed Influence: New Vulnerabilities of Structural Auxiliary Indexing in GraphRAG Pipelines with LLM}
\titlerunning{HDI: Auxiliary Structure Poisoning in GraphRAG}

\author{Jisung Park\inst{1}\href{mailto:jp407@uowmail.edu.au}{\textrm{\Letter}}\orcidID{0009-0005-8603-4125} \and 
John Le\inst{1}\orcidID{0000-0003-0019-0345} \and 
Heath Cooper\inst{1}\orcidID{0009-0001-1054-5165}}
\authorrunning{J. Park et al.}
\institute{Institute of Cybersecurity and Cryptology, University of Wollongong\\
\email{jp407@uowmail.edu.au, \{johnle, hcooper\}@uow.edu.au}}

\maketitle



\begin{abstract}
GraphRAG pipelines construct auxiliary structures during offline indexing—semantic summaries, hierarchical edges, and pre-computed scores—that determine how retrieval is prioritised at query time. Prior attacks target only instance-level components (nodes, edges, triples), overlooking these schema-level structures. We formalise Auxiliary Schema-Level Entity as a novel attack surface and propose the 3S Framework (Semantics, Structure, Scoring) for its systematic exploitation. Our Hop-Decayed Influence (HDI) attack identifies high-impact targets through query-aware influence propagation and corrupts their auxiliary structures post-indexing. Across two benchmarks (HotpotQA, 2WikiMultiHopQA) and two architectures (Microsoft GraphRAG, HippoRAG2), HDI achieves 88–94\% attack success rate while modifying as few as 0.016\% of auxiliary structures. Each modification affects up to 6.00 queries (Schema Leverage Ratio), demonstrating 1:N amplification unavailable to instance-level attacks. Manipulated structures evade perplexity and paraphrase defenses with over 99\% evasion rate, as they remain linguistically coherent system-generated artifacts. These results reveal that auxiliary schema-level entities receive implicit trust without runtime validation, constituting a structural blind spot in current GraphRAG defenses.
\url{https://github.com/Jisung-Pacific/HDI-GraphRAG-Attack}.
\end{abstract}
\keywords{GraphRAG \and Knowledge Graph Security \and Adversarial Attacks \and Schema-Level Entities}


\section{Introduction}

Large language models have achieved success \cite{brown2020languagemodelsfewshotlearners} in NLP tasks in various domains, e.g., healthcare and scientific research \cite{wu-etal-2025-medical, khodadad2025domain}.
Yet, they exhibit limitations such as hallucination, knowledge recency gaps and lack of domain specific or private knowledge \cite{kalai2025hallucinate, huang2025survey, klesel2025retrieval}.
To address these limitations, Retrieval Augmented Generation (RAG) has emerged by retrieving relevant information from external knowledge bases\cite{lewis2021RAGnlptask, gao2024retrievalaugmentedgenerationlargelanguage}. 
Existing RAG approaches have focused on semi-structured knowledge (e.g., web page, email), unstructured knowledge (e.g., textual documents), and multimodal knowledge (e.g., image, audio) \cite{cheng2025surveyknowledgeorientedretrievalaugmentedgeneration}. However, these RAG approaches still suffer from limitations such as weak structured, relational and logical knowledge representation, multi-hop traversal challenges~\cite{kalutharage2024neurosymbolic}.
\begin{figure}[htbp]
    \centering
    \includegraphics[width=\columnwidth]{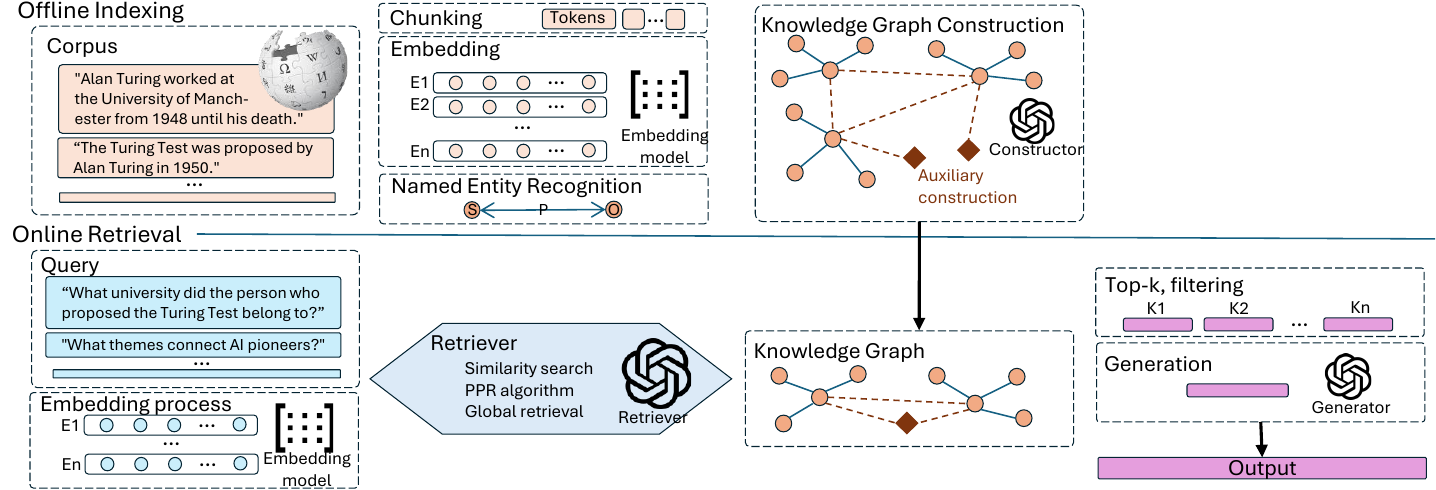}
    \caption{A process of GraphRAG pipeline}
    \label{fig:GraphRAG}
\end{figure}
KG RAG is a RAG technique that operates on knowledge graphs, leveraging structured relational knowledge where query processors, retrievers, and graph granularity interact to enhance retrieval precision\cite{han2025retrievalaugmentedgenerationgraphsgraphrag, Pan_2024Unifying, he2024gretriever, li2025simpleeffectiverolesgraphs}. GraphRAG pipeline in this research extends beyond the conventional scope of online retrieval, which merely involves interactions among queries, retrievers, and static knowledge graphs. Instead, it encompasses the complete end to end process including offline indexing, which performs knowledge extraction from raw text corpora to knowledge graph construction\cite{peng2024graph, zhang2025surveygraphretrievalaugmentedgeneration, luo2025graph}.
These pipelines construct diverse and unique auxiliary structures. These auxiliary structures either enable additional functionalities such as sense-making and multi-hop reasoning, or enhance retrieval performance efficiency and effectiveness through mechanisms like context richness and pre-computed scoring \cite{edge2025localglobalgraphrag, gutiérrez2025hipporag}. 
This approach is increasingly adopted in both academic and industrial fields across diverse domains \cite{song2025dementiaplanagentbasedframeworkmultiknowledge, wu-etal-2025-medical}.

Recent security research has revealed vulnerabilities in RAG and GraphRAG systems through systematic poisoning attacks targeting knowledge retrieval and representation. Direct graph attacks manipulate the structural components of knowledge graphs corrupting nodes, edges, or triples to propagate misinformation through relational pathways \cite{wen2025wordsdistortgraphsknowledge, liang2026graphrag, liu2025exposingprivacyrisksgraph}. Corpus to graph attacks exploit the end to end nature of GraphRAG pipelines, injecting adversarial passages into source corpora that propagate through the knowledge extraction workflow to corrupt the constructed graph \cite{zou2024poisonedrag, zhao2025ragsafetyexploringknowledge}. However, these attacks remain confined to the instance-level paradigm of conventional knowledge graphs, assuming that the graph itself constitutes the complete retrieval substrate. This perspective misses that end to end GraphRAG pipelines construct auxiliary structures during offline indexing that exist as separate layers beyond the core knowledge graph. These structures are architectural necessities for advanced functionalities including sense-making and multi-hop reasoning. They reside not in the data plane of individual facts but in the control plane that determines retrieval paths and prioritisation directly determining not merely 
\textit{what is retrieved}, but \textit{how it is ranked and interpreted}.

We propose \textbf{Auxiliary Schema-Level Entity} as an attack surface comprising components that exist beyond traditional graph granularities, systematised through the 3S Framework: \textit{Semantics} (summaries), \textit{Structure} (hierarchical edges), and \textit{Scoring} (pre-computed metrics). This qualifies as an attack surface for the following reasons. These structures are required components for advanced functionalities including sense-making and multi-hop reasoning, residing in the control plane that determines retrieval paths and prioritisation. As system generated artifacts, they are granted implicit trust and excluded from runtime validation. A single schema at hierarchical aggregation points, when manipulated, simultaneously affects numerous subordinate entities and queries providing asymmetric 1:N leverage. Even when instance-level facts remain accurate, higher level semantic abstractions dominate the interpretive context delivered to the LLM, distorting final generation. Existing defenses target anomaly detection at the node, edge, and triple level, leaving schema-level manipulation as a structural blind spot; As derived artifacts synthesised from multiple sources, they lack singular ground truth, rendering tampering verification intractable.

To demonstrate this attack surface, we suggest the \textbf{\attackname}, which targets the auxiliary extension structures Semantics, Structure, Scoring beyond instance level entities that GraphRAG pipelines uniquely employ for additional functions and efficient retrieval strategies. The attack is deployed after knowledge construction through offline indexing in the GraphRAG pipeline, but before online retrieval commences. The attack first identifies targets using type predicted queries, then applies the selected 3S attack method.

To investigate these vulnerabilities, we address the following research questions:
\noindent\textbf{RQ1:} \RQone ,
\textbf{RQ2:} \RQtwo ,
\textbf{RQ3:} \RQthree

Based on query types, we select attack targets through either scoring exploration, then manipulate the attack surfaces including semantics, structure, and scoring mechanisms. Our method achieved over 85\% attack success rate across all configurations with minimal structural perturbation. Defense evasion rates exceeded 99\% against existing detection mechanisms\cite{jain2023baselinedefensesadversarialattacks}.

Our key contributions to the understanding and advancement of GraphRAG pipeline are:

\begin{itemize}
    \item We formalise \textit{\attacksurface} (semantic summaries, hierarchical edges, pre-computed scores) as a novel attack surface in GraphRAG pipelines.
    \item We propose Hop-Decayed Influence (HDI), a gray-box attack achieving high ASR with minimal structural perturbation and 1:N leverage.
    \item We analyse defense limitations, revealing conventional instance-level defenses are insufficient for schema-level threats.
\end{itemize}
The paper proceeds as follows: Section~\ref{sec:background} covers background, Section~\ref{sec:methodology} presents methodology, Section~\ref{sec:setup} details experimental setup, Section~\ref{sec:results} reports results, and Section~\ref{sec:discussion} discusses defenses before concluding.


\section{Background}\label{sec:background}

\subsection{GraphRAG Pipeline}
GraphRAG pipeline consists of two primary phases: offline indexing and online retrieval\cite{peng2024graph, zhang2025surveygraphretrievalaugmentedgeneration}. \textbf{Offline Indexing.} Given a raw text document corpus $\mathcal{D}$, the system performs text chunking to divide documents into overlapping segments that maintain semantic continuity. Each chunk is transformed into dense vector representations via embedding functions, enabling similarity based retrieval. Concurrently, a knowledge extraction process identifies entities $\mathcal{E}$ and relations $\mathcal{R}$ using LLMs or NLP tools, representing each fact as a triple $(e_i, r_k, e_j)$. The extracted information is structured into a knowledge graph $\mathcal{G} = (\mathcal{V}, \mathcal{E}')$, where vertices include both entities and text chunks, and edges represent relationships between them. An auxiliary structure generation function $f_{\text{aux}}: \mathcal{G} \rightarrow \mathcal{A}$ creates pipeline specific schema-level information that distinguishes end to end GraphRAG from simple KG based retrieval. \textbf{Online Retrieval.} Given a query $q$, the system embeds it into the same vector space and performs similarity based search to extract relevant context. The retrieval granularity $g \in \{\text{node}, \text{edge}, \text{path}, \text{subgraph}\}$ determines the structural form of retrieved knowledge. Retrieved elements undergo ranking through scoring functions to select the top-$k$ most relevant components within context window constraints. Finally, the LLM generates answers by processing the selected context alongside the original query.

\subsection{GraphRAG Attack Methods}
Prior GraphRAG attacks operate at the instance level. \textbf{GRAGPOISON} targets relations by identifying shared relations across queries and injecting competing relations with temporal ordering~\cite{liang2026graphrag}. Corpus-level attacks exploit GraphRAG's dependency on source text: \textbf{TKPA} rewrites vulnerable text chunks to control specific query outputs, while \textbf{UKPA} perturbs coreference signals to fragment graph structure globally~\cite{wen2025wordsdistortgraphsknowledge}. \textbf{PoisonedRAG} injects malicious texts optimised to maximise retrieval similarity while inducing adversarial answers~\cite{zou2024poisonedrag}.

\section{Methodology}\label{sec:methodology}

\subsection{Threat model}
\begin{figure*}[t]
    \centering
    \includegraphics[width=\textwidth]{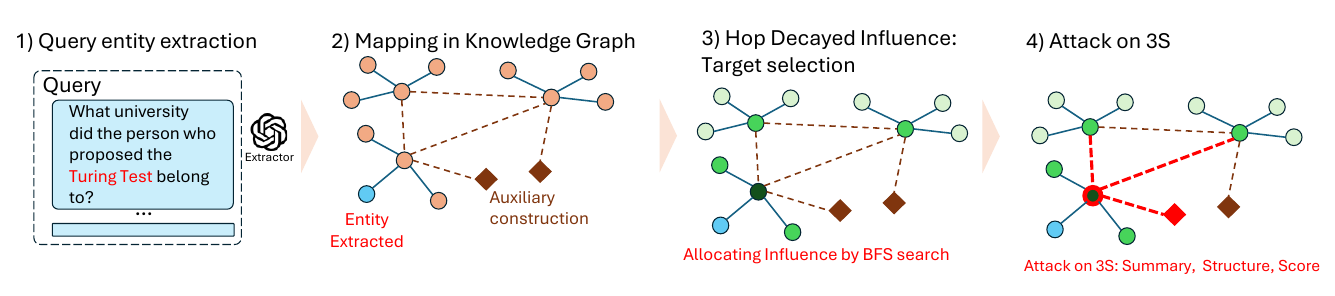}
    \caption{The process of \attackname.}
    \label{fig:Threatmodel}
\end{figure*}

\paragraph{Adversary's Objectives.} The adversary aims to degrade GraphRAG retrieval performance by corrupting auxiliary structures used for retrieval optimization. Given a set of target queries, the adversary seeks to inject minimal perturbations such that the system generates incorrect responses while bypassing existing defense mechanisms. \textbf{Attack Surface: \attacksurface.} 
We introduce \textit{\attacksurface} as an attack surface in GraphRAG systems, defined as structural components constructed during offline indexing that exist beyond instance-level elements (nodes, edges, triples). These structures augment LLM context, optimise retrieval through precomputation, and guide graph navigation; we categorise them into \textit{Semantics}, \textit{Structure}, and \textit{Scoring} (3S). Unlike prior attacks that alter graph content, our attack manipulates how the graph is traversed.
\paragraph{Adversary's Capabilities.}
The adversary can modify auxiliary structures in the constructed knowledge graph, operating post-indexing. Such capabilities arise in practical settings such as insider threats with database write access or compromised graph database interfaces. In practice, representative GraphRAG systems store auxiliary structures (e.g., summaries as plaintext Parquet files, scores in unprotected caches) without integrity verification separate from core graph data. However, write access alone is insufficient; identifying which minimal subset to corrupt for maximum cross-query impact is the core challenge. The adversary cannot access the original text corpus, cannot retrigger the indexing pipeline, and cannot control any parameters of the retriever or generator. 
\paragraph{Adversary's Knowledge.} We assume a gray-box setting where the adversary has access to a representative target query set, reflecting scenarios with partially predictable query patterns (e.g., enterprise knowledge bases with recurring analytical queries), and can query the knowledge graph through standard graph database interfaces.

\subsection{Problem Formulation}

A GraphRAG system constructs a knowledge graph from a text corpus, along with auxiliary structures for retrieval optimisation. These auxiliary structures include semantic summaries, hierarchical edges, and pre-computed scores.

Given a query set $\mathcal{Q}$, the attacker's objective is to find minimal perturbations to auxiliary structures such that retrieval fails for the maximum number of queries, subject to budget constraint $\epsilon$:

\begin{equation}
\max_{\Delta\mathcal{A}} \sum_{q \in \mathcal{Q}} \mathbb{1}\left[ \mathcal{R}(q, \mathcal{A} + \Delta\mathcal{A}) \cap \mathcal{R}(q, \mathcal{A}) = \emptyset \right]
\end{equation}

where $\mathcal{R}(q, \mathcal{A})$ denotes the retrieved context for query $q$.

\subsection{Target Selection: \attackname}

Effective target selection is essential for maximising attack impact under limited budget. We propose \attackname{} (HDI), a query aware method that identifies high influence nodes.

For each query $q$, we extract seed entities and map them to graph nodes. From each seed $s$, influence propagates with exponential decay based on shortest path distance $d(s,v)$:

\begin{equation}
\text{Influence}(v) = \sum_{q \in \mathcal{Q}} \sum_{s \in \text{Seeds}(q)} \lambda^{d(s,v)}
\end{equation}

where $\lambda \in (0,1)$ is the decay factor. Nodes beyond $h_{\max}$ hops receive zero influence. Given budget $k$, we select the top-$k$ nodes with highest cumulative influence as attack targets.

\subsection{Attack Methods: 3S Framework}
We propose the 3S attack framework targeting three categories of auxiliary structures.

\paragraph{Semantic Attack.}
Corrupts textual summaries associated with target nodes by injecting adversarial text $\delta$ that introduces false entity associations while maintaining low perplexity to evade detection.

\paragraph{Structure Attack.}
Injects misleading edges between target nodes and semantically distant nodes. For each target $v^*$, we add edges to nodes with similarity below threshold $\tau$, i.e., $\mathcal{E}' = \mathcal{E} \cup \{(v^*, u) : \text{sim}(v^*, u) < \tau \}$, creating false traversal paths.

\paragraph{Score Attack.}
Manipulates pre-computed scores to disrupt ranking: for target set $V^* = \arg\max_{v \in \mathcal{V}, |V^*|=k} \sigma(v)$, scores are demoted via $\sigma'(v^*) = \mu$, while median-ranked nodes $V_{\text{med}}$ are promoted via $\sigma'(v) = \sigma_{\text{max}}$, causing retrievers to prioritise irrelevant nodes.

\subsection{Algorithm Overview}
\begin{algorithm}[t]
\caption{Hop Decayed Influence Target Selection}
\label{alg:hdi}
\KwIn{Query set $\mathcal{Q}$, Knowledge graph $\mathcal{G}$, Decay factor $\lambda$, Max hops $h_{\max}$, Budget $k$}
\KwOut{Target node set $\mathcal{T}$}

\textbf{Initialise:} $\text{Influence}[v] \leftarrow 0$ for all $v \in \mathcal{V}$\;

\ForEach{query $q \in \mathcal{Q}$}{
    $\text{Seeds}(q) \leftarrow \textsc{ExtractEntities}(q)$\;
    \ForEach{seed $s \in \text{Seeds}(q)$}{
        $s_v \leftarrow \textsc{MapToNode}(s, \mathcal{G})$\;
        \If{$s_v \neq \text{null}$}{
            \ForEach{node $v \in \mathcal{V}$}{
                $d \leftarrow \textsc{ShortestPath}(s_v, v, \mathcal{G})$\;
                \If{$d \leq h_{\max}$}{
                    $\text{Influence}[v] \leftarrow \text{Influence}[v] + \lambda^{d}$\;
                }
            }
        }
    }
}

$\mathcal{T} \leftarrow \textsc{TopK}(\text{Influence}, k)$\;
\Return{$\mathcal{T}$}
\end{algorithm}
The attack proceeds in two phases. First, HDI computes influence scores by extracting seeds from each query and propagating influence with decay $\lambda$. The top-$k$ nodes are selected as targets. Second, the selected attack method (Semantic, Structure, or Score) is applied to each target node. The choice depends on the target pipeline's retrieval mechanism: semantic attacks are effective against embedding based systems, while structure attacks target graph traversal systems.

\section{Experiment setup}\label{sec:setup}
\paragraph{Benchmark datasets \& GraphRAG Baseline.}
To evaluate the effectiveness of \attackname attacks and observe attack patterns in multi-hop graph RAG task, we employ two established benchmark datasets: HotpotQA \cite{yang2018hotpotqa} and 2WikiMultiHopQA \cite{ho2020constructingmultihopqadataset} sourced from Wikipedia. Those datasets are sampled with 1,000 cases from each dataset due to computational budget constraints.
We compare \shortmethod against RAG systems: Microsoft GraphRAG\cite{edge2025localglobalgraphrag} uses hierarchical community detection where high-level summaries aggregate subordinate entities. HippoRAG2\cite{gutiérrez2025hipporag, gutiérrez2025ragmemory} relies on Personalised PageRank (PPR) with pre-computed scores for retrieval prioritisation.

\paragraph{Configuration setup.}
OpenAI's 'text-embedding-3-small' \cite{openai2024embeddings} and 'GPT-4o-mini'\cite{openai2024gpt4ocard} are employed. For KG construction we use networkx, tiktoken and datasets of huggingface. During graph construction, we apply standard quality control including entity deduplication and relation normalisation to ensure consistent baseline conditions. For HDI target selection, we set decay factor $\lambda = 0.5$, maximum hop distance $h_{\max} = 4$, random seed = 42 and manipulation intensity = {0.001\%, 0.01\%, 0.1\%}. Experiments were conducted on over 32GB of RAM.

\paragraph{Metric.}
To evaluate attack performance, we propose new metrics:
\subsubsection{Schema Leverage Ratio (SLR)}
The expected number of queries affected per single auxiliary structure modification.

\begin{equation}
\text{SLR} = \frac{|\mathcal{Q}_{\text{affected}}|}{|\Delta\mathcal{A}|}
\end{equation}

\noindent where $\mathcal{Q}_{\text{affected}}$ is the set of queries with altered responses after attack, and $\Delta\mathcal{A}$ is the set of modified auxiliary structures. Instance-level attacks achieve $\text{SLR} \approx 1$ as each modification targets one query. Schema-level attacks exploit shared auxiliary structures across retrieval paths, enabling $\text{SLR} \gg 1$. Higher SLR indicates greater architectural vulnerability to schema-level manipulation.


We also employ standard metrics to evaluate attack effectiveness. \textbf{Attack Success Rate (ASR)} quantifies the proportion of queries where the adversarial attack successfully manipulates the system's response to match the adversarial objective, computed as $\text{ASR} = \frac{1}{N} \sum_{i=1}^{N} \mathbf{1}[\text{attack}_i = \text{success}]$, where $N$ denotes the total number of target queries \cite{Muhammad2023QASiNa, huang2023catast}. Specifically, an attack is considered successful when a query that was correctly answered before the attack produces an incorrect answer after the attack. \textbf{F1 Score} measures token-level overlap between predicted answer $y_i$ and ground truth $y_i^*$ using the harmonic mean of precision $P_i = \frac{|\text{tokens}(y_i) \cap \text{tokens}(y_i^*)|}{|\text{tokens}(y_i)|}$ and recall $R_i = \frac{|\text{tokens}(y_i) \cap \text{tokens}(y_i^*)|}{|\text{tokens}(y_i^*)|}$, averaged across all queries as $\text{F1} = \frac{1}{N} \sum_{i=1}^{N} \frac{2 \cdot P_i \cdot R_i}{P_i + R_i}$. \textbf{Exact Match (EM)} provides strict evaluation by measuring whether normalised predicted answers exactly match ground truth: $\text{EM} = \frac{1}{|Q|} \sum_{i=1}^{|Q|} \mathbf{1}[\text{norm}(\hat{y}_i) = \text{norm}(y_i^{\text{gold}})]$, where $\text{norm}(\cdot)$ applies lowercase conversion, whitespace removal, and punctuation handling\cite{yani2021challenges}. Based on the stealthiness standard\cite{panebianco2025stealthy}, PF and PD are measured.


\section{Experimental Results}\label{sec:results}

\subsection{\textbf{RQ1:} \RQone}

\begin{table*}[!t]
    \centering
    \small
    \caption{Performance Comparison Before and After Attack Across Different Models and Datasets}
    \label{tab:combined_comparison}
    \renewcommand{\arraystretch}{1.1}
    \begin{tabular}{lcccc}
        \toprule
        & \multicolumn{2}{c}{\textbf{MSGraphRAG}} & \multicolumn{2}{c}{\textbf{HippoRAG2}} \\
        \cmidrule(lr){2-3} \cmidrule(lr){4-5}
        & HotpotQA & 2WikiMultiHopQA & HotpotQA & 2WikiMultiHopQA \\
        \midrule
        Best Config             & HDI-T         & HDI-T            & HDI-T         & HDI-C         \\
        Our ASR (\%) & \textbf{91.69} & \textbf{88.17} & \textbf{88.24} & \textbf{94.44} \\
        PoisonedRAG (\%) & 80.69  & 80.42 & 81.49 & 85.89    \\
        Gragpoison (\%) & 85.64  & 88.68 & 86.28 & 87.24    \\
        \bottomrule
    \end{tabular}
    \vspace{-0.1in}
\end{table*}

Table~\ref{tab:combined_comparison} shows attack effectiveness across configurations. Microsoft GraphRAG on HotpotQA achieves 91.69\% ASR, while 2WikiMultiHopQA reaches 88.17\%. HippoRAG2 shows similar vulnerability: 88.24\% ASR on HotpotQA and 94.44\% on 2WikiMultiHopQA, the highest across all configurations. The two systems differ in vulnerability patterns. Microsoft GraphRAG's hierarchical community structure amplifies attack impact: corrupting one high-level summary affects all queries routed through that community. HippoRAG2's Personalised PageRank distributes retrieval across multiple paths, offering some resilience. Still, ASR exceeds 86\% in every configuration. Schema-level entities are exploitable regardless of architectural choices.


\subsection{\textbf{RQ2:} \RQtwo}

\begin{table*}[!t]
    \centering
    \small
    \caption{Attack Efficiency Comparison Across Different GraphRAG Models and Datasets. \textit{HDI-T: HDI with Structure attack; HDI-C: HDI with Score attack; HDI-S: HDI with Semantic attack. SLR = Schema Leverage Ratio (affected queries / modifications). Mod. Ratio(\%) = \# Modified / Total Structures $\times$ 100.}}
    \label{tab:efficiency_comparison}
    \renewcommand{\arraystretch}{1.1}
    \begin{tabular}{lcccc}
        \toprule
        & \multicolumn{2}{c}{\textbf{MSGraphRAG}} & \multicolumn{2}{c}{\textbf{HippoRAG2}} \\
        \cmidrule(lr){2-3} \cmidrule(lr){4-5}
        & HotpotQA & 2WikiMultiHopQA & HotpotQA & 2WikiMultiHopQA \\
        \midrule
        Best Config   & HDI-T    & HDI-T    & HDI-T   & HDI-C  \\
        \midrule
        Total Structures  & 6,328  & 9,328  & 440,020  & 311,419  \\
        \# Modified       & 300       & 450     & 330    & 49     \\
        Mod. Ratio (\%)   & 4.74      & 4.82    & 0.075   & 0.016  \\
        \midrule
        \textbf{Our SLR}   & 4.57 & 3.41    & 5.80   & 6.00         \\
        \textbf{PoisonedRag's SLR}  & 1.36 & 1.25  & 0.42    & 0.32         \\
        \bottomrule
    \end{tabular}
\end{table*}

Table~\ref{tab:efficiency_comparison} quantifies attack footprint. HDI achieves high success rates with minimal modifications.

HippoRAG2 on 2WikiMultiHopQA requires only 49 modified structures out of 311,419 (0.016\%) of the total that yet reaches 94.44\% ASR. On HotpotQA, 330 modifications (0.075\%) produce 88.24\% ASR. Microsoft GraphRAG has fewer total structures (6,328 on HotpotQA, 9,328 on 2WikiMultiHopQA), so modification ratios are higher (4.74\% and 4.82\%), but absolute counts remain low: 300–450 modifications.

Schema Leverage Ratio (SLR) measures queries affected per modification. Our approach achieves SLR of 3.41–6.00, compared to PoisonedRAG's 0.32–1.36. The gap is largest on HippoRAG2-2WikiMultiHopQA: our SLR (6.00) versus PoisonedRAG's (0.32). This 1:N leverage that one schema modification affecting many queries is unavailable to instance-level attacks.

Attack type matters. We evaluate defense evasion using each configuration's best performing attack type (Table~\ref{tab:defense_evasion}). HippoRAG2 prefers Structure attacks on HotpotQA but Score attacks (C) on 2WikiMultiHopQA, reflecting PPR's sensitivity to pre-computed rankings.


\subsection{\textbf{RQ3:} \RQthree}

\begin{table*}[!t]
    \centering
    \small
    \caption{Defense Evasion Performance (RQ3): Perplexity Filter and Paraphrase Defense}
    \label{tab:defense_evasion}
    \renewcommand{\arraystretch}{1.1}
    \begin{tabular}{lcccc}
        \toprule
        & \multicolumn{2}{c}{\textbf{MSGraphRAG}} & \multicolumn{2}{c}{\textbf{HippoRAG2}} \\
        \cmidrule(lr){2-3} \cmidrule(lr){4-5}
        & HotpotQA & 2WikiMultiHopQA & HotpotQA & 2WikiMultiHopQA \\
        \midrule
        Best Type        & HDI-T & HDI-T & HDI-T & HDI-C \\
        \midrule
        Perplexity Evasion (\%)        & 99.85 & 99.50 & 99.87           & 99.79 \\
        Paraphrase Evasion (\%) & 99.13 & 99.14 & 99.72           & 99.27 \\
        Detection Rate (\%)     & 0.07           & 0.09 & 0.05            & 0.08           \\
        \bottomrule
    \end{tabular}
\end{table*}

Table~\ref{tab:defense_evasion} evaluates HDI against perplexity filtering and paraphrase defense. Structure attacks (T) serve as the test configuration.

Detection rates are negligible. Perplexity filtering misses 99.50\% – 99.87\% of modifications. Paraphrase defense fails similarly, with evasion rates of 99.13\% – 99.72\%. Across all configurations, detection stays below 0.1\%.

These defenses target instance-level anomalies: odd entity names, implausible triples, statistically unusual text. Schema-level manipulations evade them because corrupted summaries are still LLM generated and linguistically fluent. Injected edges connect semantically plausible entity pairs. Numerical scores bypass text based detection entirely.

Auxiliary structures receive implicit trust. Auxiliary structures, being system generated artifacts, receive no runtime validation. Current defenses assume attacks manifest as anomalous content. Schema-level attacks produce normal looking content through abnormal selection and aggregation. This mismatch leaves GraphRAG systems exposed.

\subsection{Ablation Study}
\begin{figure*}[t]
    \centering
    \includegraphics[width=\textwidth]{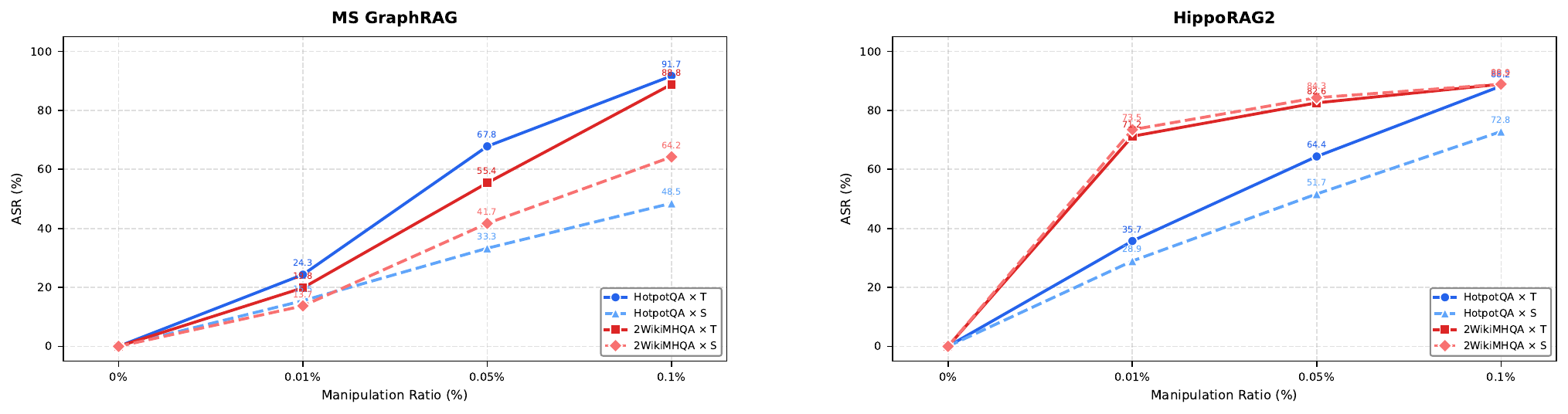}
    \caption{ASR scaling with modification intensity across GraphRAG systems. Left: MS GraphRAG. Right: HippoRAG2. X-axis: proportion of modification intensity (\%). Y-axis: Attack Success Rate (\%). T: Structure attack. S: Semantic attack.}
    \label{fig:ablation}
\end{figure*}

Modification intensity Scaling Analysis. Figure~\ref{fig:ablation} shows ASR scaling with the proportion of modified structures. Structure attacks (T) consistently outperform Semantic attacks (S) on MS GraphRAG, achieving 91.69\% versus 48.52\% ASR when modifying 0.1\% of total structures on HotpotQA. The two systems exhibit distinct patterns: MS GraphRAG shows near-linear growth, while HippoRAG2 on 2WikiMultiHopQA demonstrates early saturation, reaching 71–73\% ASR by modifying merely 0.01\% of structures. This indicates that densely connected auxiliary structures enable high attack impact with minimal perturbation. Notably, on HippoRAG2-2WikiMultiHopQA, both attack methods converge to 88.89\% ASR, though Semantic attacks achieve superior efficiency.


\section{Discussion \& Defense}\label{sec:discussion}

Prior stealthiness research in adversarial RAG has focused on detecting statistical anomalies at the instance level perplexity filtering and paraphrase detection assuming adversarial content manifests as linguistic outliers~\cite{jain2023baselinedefensesadversarialattacks, panebianco2025stealthy}. Schema-level manipulations evade these defenses for two reasons: auxiliary structures are LLM generated and thus linguistically fluent by construction, and aggregated summaries lack singular ground truth against which tampering can be verified.

We identify three potential defense directions. First, \textit{provenance tracking} maintains lineage between schema entities and their source instances, enabling consistency audits. Second, \textit{runtime verification} detects semantic drift between retrieved summaries and underlying nodes at query time. Third, \textit{robust aggregation} applies outlier resistant functions to summarisation and scoring mechanisms. However, these defenses impose computational overhead that partially negates the efficiency gains schema-level structures were designed to provide, presenting an inherent trade-off between retrieval performance and adversarial robustness.

\section{Ethical consideration} All experiments were conducted on public benchmark datasets and open source implementations in isolated environments. No attacks targeted production systems. We follow responsible disclosure practices, sharing findings with affected project maintainers prior to publication. This research aims to improve GraphRAG security by enabling developers to understand and mitigate schema-level vulnerabilities.


\section{Conclusion \& Future work}\label{sec:conclusion}

This work identifies Auxiliary Schema-Level Entity as a distinct attack surface in GraphRAG pipelines and demonstrates through the 3S Framework and Hop-Decayed Influence attack that minimal perturbation yields high attack success rates with 1:N amplification unavailable to instance-level approaches. These findings expose that auxiliary structures receive implicit trust without runtime validation, rendering current defenses inadequate for schema-level threats.

Future research will develop provenance-aware defense mechanisms that maintain lineage between schema-level entities and source instances, enabling consistency verification through cross referencing aggregated summaries against underlying nodes. We plan to investigate robust aggregation functions resistant to outlier manipulation and extend HDI to recently proposed GraphRAG pipeline with temporal and multi-modal retrieval capabilities. Additionally, developing standardised benchmark suites for evaluating schema-level attack resilience will enable systematic comparison of defense strategies across diverse GraphRAG implementations.

Further evaluation should extend to other GraphRAG applications (e.g., summarisation, decision support) and out-of-distribution queries. Key limitations include: the post-indexing write access assumption may not hold in all deployments, defense evaluation is limited to text-level detection methods mismatched against score and structure tampering, and generalisation across backbone LLMs remains unverified.


\bibliographystyle{splncs04}

\bibliography{referencelist}

@misc{kalai2025hallucinate,
      title={{Why Language Models Hallucinate}}, 
      author={Adam Tauman Kalai and Ofir Nachum and Santosh S. Vempala and Edwin Zhang},
      year={2025},
      month={sep},
}

@inproceedings{brown2020languagemodelsfewshotlearners,
      title={Language Models are Few-Shot Learners}, 
      author={Tom B. Brown and Benjamin Mann and Nick Ryder and Melanie Subbiah and Jared Kaplan and Prafulla Dhariwal and Arvind Neelakantan and Pranav Shyam, et al.},
      booktitle={Advances in Neural Information Processing Systems},
      volume={33},
      pages={1877--1901},
      year={2020},
      publisher={Curran Associates, Inc.}
}

@article{huang2025survey,
  title={A survey on hallucination in large language models: Principles, taxonomy, challenges, and open questions},
  author={Huang, Lei and Yu, Weijiang and Ma, Weitao and Zhong, Weihong and Feng, Zhangyin and Wang, Haotian and Chen, Qianglong and Peng, Weihua and Feng, Xiaocheng and Qin, Bing and others},
  journal={ACM Transactions on Information Systems},
  volume={43},
  number={2},
  pages={1--55},
  year={2025},
  publisher={ACM New York, NY}
}

@article{klesel2025retrieval,
  title={Retrieval-Augmented Generation (RAG) M. Klesel, HF Wittmann},
  author={Klesel, Michael and Wittmann, H Felix},
  journal={Business \& Information Systems Engineering},
  pages={1--11},
  year={2025},
  publisher={Springer}
}

@inproceedings{lewis2021RAGnlptask,
      title={Retrieval-Augmented Generation for Knowledge-Intensive NLP Tasks}, 
      author={Patrick Lewis and Ethan Perez and Aleksandra Piktus and Fabio Petroni and Vladimir Karpukhin and at al.},
      booktitle={Advances in Neural Information Processing Systems},
      volume={33},
      pages={9459--9474},
      year={2020},
      publisher={Curran Associates, Inc.}
}

@misc{gao2024retrievalaugmentedgenerationlargelanguage,
      title={Retrieval-Augmented Generation for Large Language Models: A Survey}, 
      author={Yunfan Gao and Yun Xiong and Xinyu Gao and Kangxiang Jia and Jinliu Pan and Yuxi Bi and Yi Dai and Jiawei Sun and Meng Wang and Haofen Wang},
      year={2024},
}

@misc{cheng2025surveyknowledgeorientedretrievalaugmentedgeneration,
      title={A Survey on Knowledge-Oriented Retrieval-Augmented Generation}, 
      author={Mingyue Cheng and Yucong Luo and Jie Ouyang and Qi Liu and Huijie Liu and Li Li and Shuo Yu and Bohou Zhang and Jiawei Cao and Jie Ma and Daoyu Wang and Enhong Chen},
      year={2025},
}

@inproceedings{he2024gretriever,
  title={{G-Retriever}: Retrieval-Augmented Generation for Textual Graph Understanding and Question Answering},
  author={He, Xiaoxin and Tian, Yijun and Sun, Yifei and Chawla, Nitesh V. and Laurent, Thomas and LeCun, Yann and Bresson, Xavier and Hooi, Bryan},
  booktitle={Advances in Neural Information Processing Systems (NeurIPS)},
  volume={37},
  year={2024},
  url={https://openreview.net/forum?id=MPJ3oXtTZl}
}

@article{Pan_2024Unifying,
   title={Unifying Large Language Models and Knowledge Graphs: A Roadmap},
   volume={36},
   ISSN={2326-3865},
   url={http://dx.doi.org/10.1109/TKDE.2024.3352100},
   DOI={10.1109/tkde.2024.3352100},
   number={7},
   journal={IEEE Transactions on Knowledge and Data Engineering},
   publisher={Institute of Electrical and Electronics Engineers (IEEE)},
   author={Pan, Shirui and Luo, Linhao and Wang, Yufei and Chen, Chen and Wang, Jiapu and Wu, Xindong},
   year={2024},
   month=jul, pages={3580–3599} }

@article{peng2024graph,
  title={Graph retrieval-augmented generation: A survey},
  author={Peng, Boci and Zhu, Yun and Liu, Yongchao and Bo, Xiaohe and Shi, Haizhou and Hong, Chuntao and Zhang, Yan and Tang, Siliang},
  year={2024}
}

@misc{zhang2025surveygraphretrievalaugmentedgeneration,
      title={A Survey of Graph Retrieval-Augmented Generation for Customized Large Language Models}, 
      author={Qinggang Zhang and Shengyuan Chen and Yuanchen Bei and Zheng Yuan and Huachi Zhou and Zijin Hong and Hao Chen and Yilin Xiao and Chuang Zhou and Junnan Dong and Yi Chang and Xiao Huang},
      year={2025},
}

@mastersthesis{khodadad2025domain,
  title={Domain-Specific Adaptation and Multi-Hop Reasoning in Chemistry and Biomedicine},
  author={Khodadad, Mohammad},
  year={2025},
  school={McMaster University},
  address={Hamilton, Ontario, Canada},
  type={{MASc} Thesis},
  note={Advisor: Hamidreza Mahyar}
}

@misc{han2025retrievalaugmentedgenerationgraphsgraphrag,
      title={Retrieval-Augmented Generation with Graphs (GraphRAG)}, 
      author={Haoyu Han and Yu Wang and Harry Shomer and Kai Guo and Jiayuan Ding and Yongjia Lei and Mahantesh Halappanavar and Ryan A. Rossi and Subhabrata Mukherjee and Xianfeng Tang and Qi He and Zhigang Hua and Bo Long and Tong Zhao and Neil Shah and Amin Javari and Yinglong Xia and Jiliang Tang},
      year={2025},
}

@misc{li2025simpleeffectiverolesgraphs,
      title={Simple Is Effective: The Roles of Graphs and Large Language Models in Knowledge-Graph-Based Retrieval-Augmented Generation}, 
      author={Mufei Li and Siqi Miao and Pan Li},
      year={2025},
}

@inproceedings{huang2023catast,
  title={Catastrophic Jailbreak of Open-source LLMs via Exploiting Generation},
  author={Huang, Yangsibo and Gupta, Samyak and Xia, Mengzhou and Li, Kai and Chen, Danqi},
  booktitle={International Conference on Learning Representations},
  year={2024},
}

@inproceedings{zou2024poisonedrag,
  title={{PoisonedRAG}: Knowledge Corruption Attacks to Retrieval-Augmented Generation of Large Language Models},
  author={Zou, Wei and Geng, Runpeng and Wang, Binghui and Jia, Jinyuan},
  booktitle={34th USENIX Security Symposium (USENIX Security 25)},
  pages={3827--3844},
  year={2025}
}

@inproceedings{liang2026graphrag,
  title     = {{GraphRAG} under Fire},
  author    = {Liang, Jiacheng and Wang, Yuhui and Li, Changjiang and Zhu, Rongyi and Jiang, Tanqiu and Gong, Neil and Wang, Ting},
  booktitle = {2026 IEEE Symposium on Security and Privacy (SP)},
  pages     = {1316--1333},
  year      = {2026},
  doi       = {10.1109/SP63933.2026.00070}
}

@misc{wen2025wordsdistortgraphsknowledge,
      title={A Few Words Can Distort Graphs: Knowledge Poisoning Attacks on Graph-based Retrieval-Augmented Generation of Large Language Models}, 
      author={Jiayi Wen and Tianxin Chen and Zhirun Zheng and Cheng Huang},
      year={2025},
}

@misc{zhao2025ragsafetyexploringknowledge,
      title={RAG Safety: Exploring Knowledge Poisoning Attacks to Retrieval-Augmented Generation}, 
      author={Tianzhe Zhao and Jiaoyan Chen and Yanchi Ru and Haiping Zhu and Nan Hu and Jun Liu and Qika Lin},
      year={2025},
}

@misc{liu2025exposingprivacyrisksgraph,
      title={Exposing Privacy Risks in Graph Retrieval-Augmented Generation}, 
      author={Jiale Liu and Jiahao Zhang and Suhang Wang},
      year={2025},
}

@misc{gutiérrez2025hipporag,
      title={HippoRAG: Neurobiologically Inspired Long-Term Memory for Large Language Models}, 
      author={Bernal Jiménez Gutiérrez and Yiheng Shu and Yu Gu and Michihiro Yasunaga and Yu Su},
      year={2025},
}

@misc{gutiérrez2025ragmemory,
      title={From RAG to Memory: Non-Parametric Continual Learning for Large Language Models}, 
      author={Bernal Jiménez Gutiérrez and Yiheng Shu and Weijian Qi and Sizhe Zhou and Yu Su},
      year={2025},
}

@misc{edge2025localglobalgraphrag,
      title={From Local to Global: A Graph RAG Approach to Query-Focused Summarization}, 
      author={Darren Edge and Ha Trinh and Newman Cheng and Joshua Bradley and Alex Chao and Apurva Mody and Steven Truitt and Dasha Metropolitansky and Robert Osazuwa Ness and Jonathan Larson},
      year={2025},
}

@article{luo2025graph,
  title={Graph-r1: Towards agentic graphrag framework via end-to-end reinforcement learning},
  author={Luo, Haoran and Chen, Guanting and Lin, Qika and Guo, Yikai and Xu, Fangzhi and Kuang, Zemin and Song, Meina and Wu, Xiaobao and Zhu, Yifan and Tuan, Luu Anh and others},
  year={2025},
}

@inproceedings{yang2018hotpotqa,
  title={{HotpotQA}: A Dataset for Diverse, Explainable Multi-hop Question Answering},
  author={Yang, Zhilin and Qi, Peng and Zhang, Saizheng and Bengio, Yoshua and Cohen, William W. and Salakhutdinov, Ruslan and Manning, Christopher D.},
  booktitle={Proceedings of the 2018 Conference on Empirical Methods in Natural Language Processing},
  pages={2369--2380},
  year={2018},
  address={Brussels, Belgium},
  publisher={Association for Computational Linguistics}
}

@inproceedings{ho2020constructingmultihopqadataset,
  title={Constructing A Multi-hop {QA} Dataset for Comprehensive Evaluation of Reasoning Steps},
  author={Ho, Xanh and Duong Nguyen, Anh-Khoa and Sugawara, Saku and Aizawa, Akiko},
  booktitle={Proceedings of the 28th International Conference on Computational Linguistics},
  pages={6609--6625},
  year={2020},
  address={Barcelona, Spain (Online)},
  publisher={International Committee on Computational Linguistics},
  doi={10.18653/v1/2020.coling-main.580}
}

@misc{openai2024gpt4ocard,
      title={GPT-4o System Card}, 
      author={OpenAI and : and Aaron Hurst and Adam Lerer and Adam P. Goucher and Adam Perelman and Aditya Ramesh and Aidan Clark and AJ Ostrow and Akila Welihinda and Alan Hayes and Alec Radford and Aleksander Mądry and Alex Baker-Whitcomb, et al.},
      year={2024},
}

@misc{openai2024embeddings,
  author = {OpenAI},
  title = {New embedding models and API updates},
  year = {2024},
  howpublished = {\url{https://openai.com/index/new-embedding-models-and-api-updates/}}
}

@INPROCEEDINGS{Muhammad2023QASiNa,
  author={Rizqullah, Muhammad Razif and Purwarianti, Ayu and Aji, Alham Fikri},
  booktitle={2023 10th International Conference on Advanced Informatics: Concept, Theory and Application (ICAICTA)}, 
  title={QASiNa: Religious Domain Question Answering Using Sirah Nabawiyah}, 
  year={2023},
  volume={},
  number={},
  pages={1-6},
  doi={10.1109/ICAICTA59291.2023.10390123}}

@misc{song2025dementiaplanagentbasedframeworkmultiknowledge,
      title={DEMENTIA-PLAN: An Agent-Based Framework for Multi-Knowledge Graph Retrieval-Augmented Generation in Dementia Care}, 
      author={Yutong Song and Chenhan Lyu and Pengfei Zhang and Sabine Brunswicker and Nikil Dutt and Amir Rahmani},
      year={2025},
}

@inproceedings{wu-etal-2025-medical,
    title = "Medical Graph {RAG}: Evidence-based Medical Large Language Model via Graph Retrieval-Augmented Generation",
    author = "Wu, Junde  and
      Zhu, Jiayuan  and
      Qi, Yunli  and
      Chen, Jingkun  and
      Xu, Min  and
      Menolascina, Filippo  and
      Jin, Yueming  and
      Grau, Vicente",
    editor = "Che, Wanxiang  and
      Nabende, Joyce  and
      Shutova, Ekaterina  and
      Pilehvar, Mohammad Taher",
    booktitle = "Proceedings of the 63rd Annual Meeting of the Association for Computational Linguistics (Volume 1: Long Papers)",
    month = jul,
    year = "2025",
    address = "Vienna, Austria",
    publisher = "Association for Computational Linguistics",
    url = "https://aclanthology.org/2025.acl-long.1381/",
    doi = "10.18653/v1/2025.acl-long.1381",
    pages = "28443--28467",
    ISBN = "979-8-89176-251-0"
}

@article{yani2021challenges,
  title={Challenges, techniques, and trends of simple knowledge graph question answering: a survey},
  author={Yani, Mohammad and Krisnadhi, Adila Alfa},
  journal={Information},
  volume={12},
  number={7},
  pages={271},
  year={2021},
  publisher={MDPI}
}

@inproceedings{panebianco2025stealthy,
  author    = {Panebianco, Francesco and D'Onghia, Mario and 
               Zanero, Stefano and Carminati, Michele},
  title     = {How Stealthy is Stealthy? {S}tudying the Efficacy of 
               Black-Box Adversarial Attacks in the Real World},
  booktitle = {ICT Systems Security and Privacy Protection},
  series    = {IFIP Advances in Information and Communication Technology},
  volume    = {746},
  pages     = {140--154},
  publisher = {Springer, Cham},
  year      = {2025},
  editor    = {Nemec Zlatolas, Lili and Rannenberg, Kai and 
               Welzer, Tatjana and Garcia-Alfaro, Joaquin},
  note      = {40th IFIP International Conference, SEC 2025},
  doi       = {10.1007/978-3-031-92886-4_10}
}

@misc{jain2023baselinedefensesadversarialattacks,
      title={Baseline Defenses for Adversarial Attacks Against Aligned Language Models}, 
      author={Neel Jain and Avi Schwarzschild and Yuxin Wen and Gowthami Somepalli and John Kirchenbauer and Ping-yeh Chiang and Micah Goldblum and Aniruddha Saha and Jonas Geiping and Tom Goldstein},
      year={2023},
}

@inproceedings{kalutharage2024neurosymbolic,
  author    = {Kalutharage, Chathuranga Sampath and Liu, Xiaodong and Chrysoulas, Christos and Bamgboye, Oluwaseun},
  title     = {Neurosymbolic Learning in the {XAI} Framework for Enhanced Cyberattack Detection with Expert Knowledge Integration},
  booktitle = {ICT Systems Security and Privacy Protection (IFIP SEC 2024)},
  series    = {IFIP Advances in Information and Communication Technology},
  volume    = {710},
  pages     = {236--249},
  publisher = {Springer},
  year      = {2024},
  doi       = {10.1007/978-3-031-65175-5_17}
}

\end{document}